\documentclass{sig-alternate}
\usepackage{endnotes}
\usepackage{times}
\usepackage{balance}
\usepackage{flushend}
\usepackage{amsfonts}
\usepackage{amssymb}
\usepackage{amscd}
\usepackage{amsmath}
\usepackage{array}
\usepackage{epstopdf}
\usepackage{graphics}
\usepackage{epsfig}
\usepackage{algorithm, algorithmic}
\usepackage{multirow}
\usepackage{graphicx}
\usepackage{subfigure}
\usepackage[
  breaklinks=true,
  unicode=true,
  urlcolor = blue,
  colorlinks = true,
  citecolor = blue,
  linkcolor = blue]{hyperref}
\usepackage[hyphenbreaks]{breakurl}
\usepackage{tabularx}

\newcommand{\ignore}[1]{}
\newcommand{\revised}[1]{}
\newcommand\comment[1]{}

\usepackage{color}
\usepackage{listings}
\usepackage{soul}
\usepackage{graphicx}
\date{today}
\begin{document}

%\title{the title}
%\special{papersize=8.5in,11in}
%\setlength{\pdfpageheight}{\paperheight}
%\setlength{\pdfpagewidth}{\paperwidth}

%
% --- Author Metadata here ---
%\conferenceinfo{WOODSTOCK}{'97 El Paso, Texas USA}
%\CopyrightYear{2007} % Allows default copyright year (20XX) to be over-ridden - IF NEED BE.
%\crdata{0-12345-67-8/90/01}  % Allows default copyright data (0-89791-88-6/97/05) to be over-ridden - IF NEED BE.
% --- End of Author Metadata ---
%\titlenote{The corresponding vulnerability and attack schemes have been reported to Google security team.}
\title{Your Voice Assistant is Mine: How to Abuse Speakers to Steal Information and Control Your Phone
\titlenote{Responsible disclosure: We have reported the vulnerability of Google Search app and corresponding attack schemes to Google security team on May 16th 2014.}
\titlenote{Demo video can be found on the following website: \url{https://sites.google.com/site/demogvs/}}
}
%\subtitle{[Extended Abstract]
%\titlenote{A full version of this paper is available as
%\textit{Author's Guide to Preparing ACM SIG Proceedings Using
%\LaTeX$2_\epsilon$\ and BibTeX} at

\numberofauthors{1} %  in this sample file, there are a *total*

\author{
\alignauthor
Wenrui Diao, Xiangyu Liu, Zhe Zhou, and Kehuan Zhang\\
        \vspace{3pt}
       \affaddr{Department of Information Engineering}\\
       \affaddr{The Chinese University of Hong Kong}\\
       \vspace{3pt}
       \email{\{dw013, lx012, zz113, khzhang\}@ie.cuhk.edu.hk}
%\alignauthor
%Kehuan Zhang\\
%       \affaddr{Department of Information Engineering}\\
%       \affaddr{The Chinese University of Hong Kong}\\
%       \email{khzhang@ie.cuhk.edu.hk}
%\and
}
% There's nothing stopping you putting the seventh, eighth, etc.
% author on the opening page (as the 'third row') but we ask,
% for aesthetic reasons that you place these 'additional authors'
% in the \additional authors block, viz.
\additionalauthors{}
% Just remember to make sure that the TOTAL number of authors
% is the number that will appear on the first page PLUS the
% number that will appear in the \additionalauthors section.

%\titlebanner{banner above paper title}        % These are ignored unless
%\preprintfooter{short description of paper}   % 'preprint' option specified.
%\authorinfo{Name1}
%           {Affiliation1}
%           {Email1}

\maketitle

\begin{abstract}
    Previous research about sensor based attacks on Android platform focused mainly on accessing or controlling over sensitive device components, such as camera, microphone and GPS. These approaches get data from sensors directly and need corresponding sensor invoking permissions.

    This paper presents a novel approach (GVS-Attack) to launch permission bypassing attacks from a zero permission Android application (VoicEmployer) through the speaker. The idea of GVS-Attack utilizes an Android system built-in voice assistant module -- \underline{G}oogle \underline{V}oice \underline{S}earch. Through Android Intent mechanism, VoicEmployer triggers Google Voice Search to the foreground, and then plays prepared audio files (like ``\emph{call number 1234 5678}'') in the background. Google Voice Search can recognize this voice command and execute corresponding operations. With ingenious designs, our GVS-Attack can forge SMS/Email, access privacy information, transmit sensitive data and achieve remote control without any permission.

    Also we found a vulnerability of status checking in Google Search app, which can be utilized by GVS-Attack to dial arbitrary numbers even when the phone is securely locked with password. A prototype of VoicEmployer has been implemented to demonstrate the feasibility of GVS-Attack in real world. In theory, nearly all Android devices equipped with Google Services Framework can be affected by GVS-Attack. This study may inspire application developers and researchers rethink that zero permission doesn't mean safety and the speaker can be treated as a new attack surface.

\end{abstract}

%\category{K.6.5}{Security and Protection}{Unauthorized access}[]
%\category{I.2.7}{Natural Language Processing}{Speech recognition and synthesis}[]
%
%\terms{Security}
%
%\keywords{Android, Speaker, Google Voice Search, Speech Recognition, Privacy Stealing, Remote Voice Control}

\section{Introduction}
\label{attest:intro}
In recent years, smartphones are becoming more and more popular, among which Android OS pushed past 80\% market share~\cite{android_market_share}. One attraction of smartphones is that users can install applications (\emph{apps} for short) by themselves conveniently. But this convenience also brings serious malicious application problems which have been noticed by both academic and industry areas. According to Kaspersky's annual security report~\cite{kasperskyreport2013}, Android platform attracted a whopping 98.05\% of known malware in 2013.

%According to F-Secure's threat report~\cite{fsecure_thread_report}, 97\% of all new mobile malware was targeting Android devices in 2013.

Current Android phones are equipped with several kinds of sensors, such as light sensor, accelerometer, microphone, GPS, etc. They enhance the user experience and can be used to develop creative apps. However, these sensors also could be utilized as powerful weapons for mobile malwares to steal user privacy. Taking microphone as an example, a malicious app can record phone conversations which may contain sensitive business information directly. With some special designs, even credit card and PIN number can be extracted from recorded voice~\cite{schlegel2011soundcomber}.

Nearly all previous sensor based attacks~\cite{schlegel2011soundcomber, owusu2012accessory, cai2011touchlogger, aviv2012practicality, simon2013pin, placeraider, hasan2013sensing} only considered invoking input type of device components to achieve malicious targets, such as accelerometer for device posture analysis and microphone for audio recording. These attacks are based on accessing or controlling over sensitive sensors directly, which means specific permissions are needed, such as \texttt{CAMERA} for camera, \texttt{RECORD\_AUDIO} for microphone, and \texttt{ACCESS\_FINE\_LOCATION} for GPS. Actually output type of permission-free device components (e.g. speaker) also can be utilized to launch attacks, namely an indirect approach.

\vspace{3pt}
Consider the following question:\\
\fbox{
  \parbox{0.45\textwidth}{
\emph{Q: To a zero permission Android app which only invokes permission-free sensors, what malicious targets it can achieve?}
  }
}

\vspace{3pt}
In general, zero permission means harmlessness and extremely limited functions. However, our research results show:\\
\fbox{
  \parbox{0.45\textwidth}{
\emph{A: Through invoking the speaker, this zero permission app can make phone calls, forge SMS/Email, steal personal schedules, get the user location, transmit data remotely, etc.}
  }
}

\vspace{3pt}
This paper presents a novel attack scheme based on the speaker and an Android built-in voice assistant module -- Google Voice Search (so we call it GVS-Attack). GVS-Attack can be launched through a totally \textbf{zero permission} Android malware (called VoicEmployer correspondingly). This attack can achieve varied malicious targets based on sensitive permissions bypassing.

\begin{figure}[htb]
  \centering
    \subfigure[Voice Dialer Mode]{
    \label{fig:dial_mode} %% label for second subfigure
    \includegraphics[width=3.75cm]{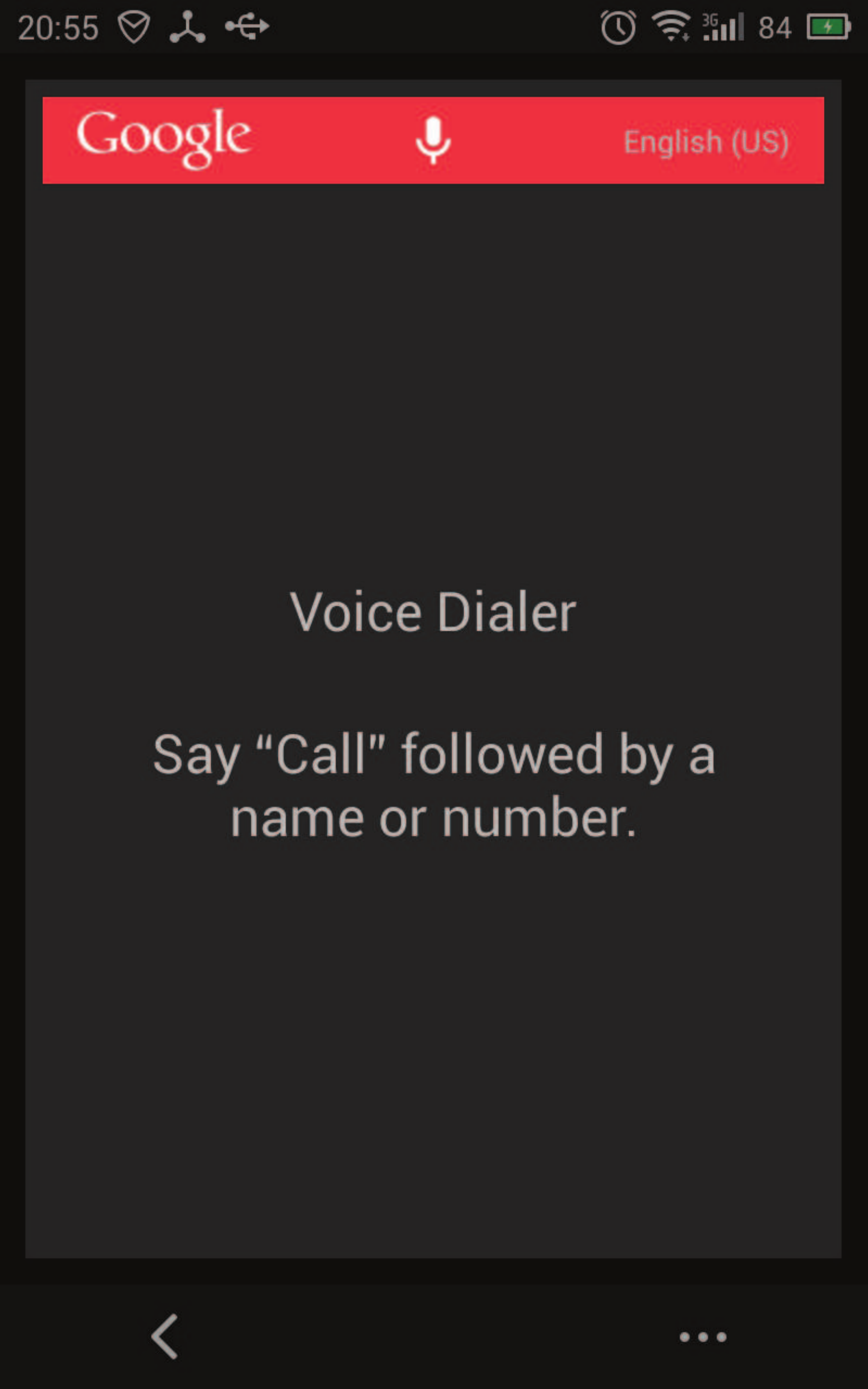}}
  \quad \quad
  \subfigure[Velvet Mode]{
    \label{fig:full_mode} %% label for first subfigure
    \includegraphics[width=3.75cm]{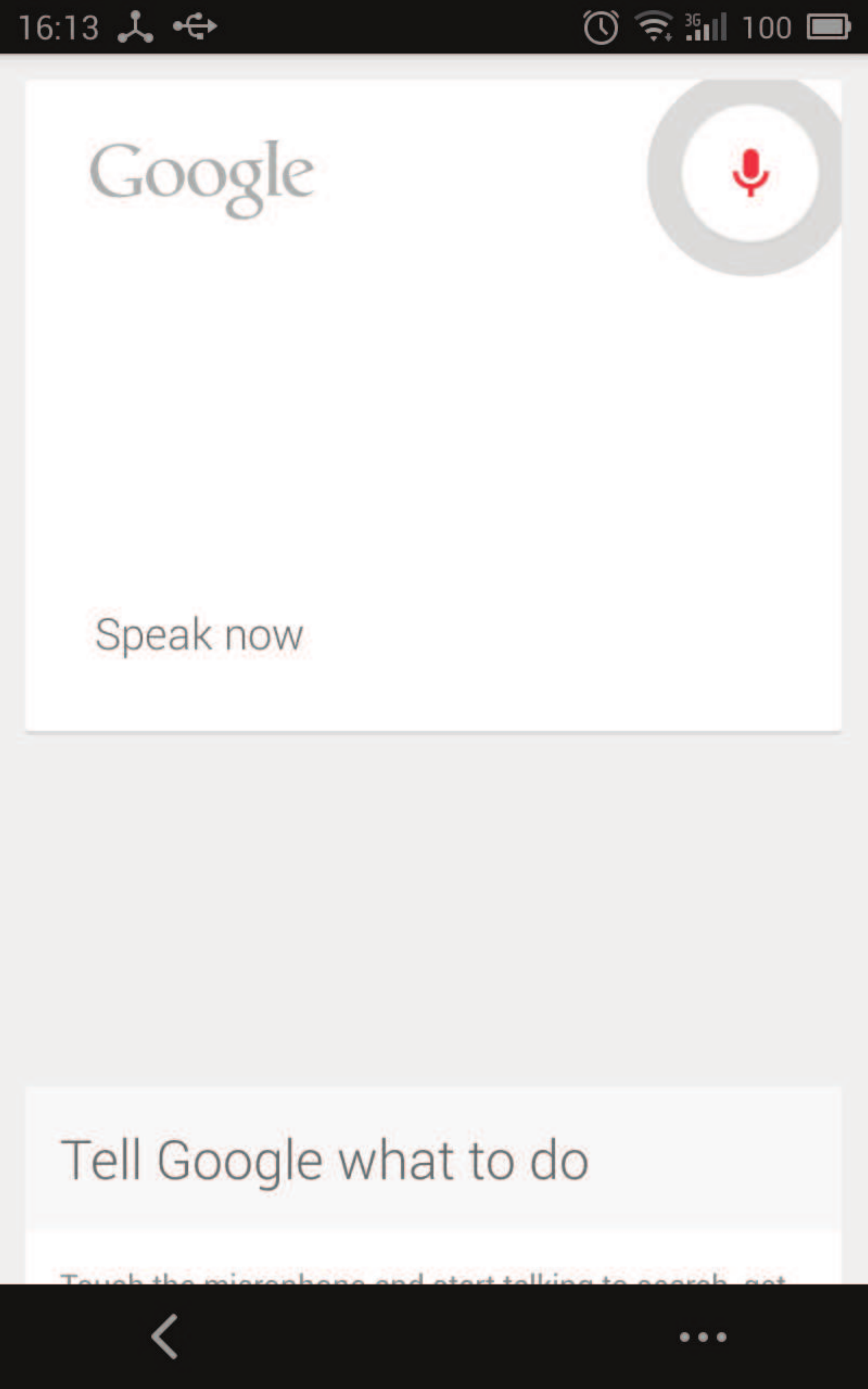}}
%  \vspace{-5pt}
  \caption{Google Voice Search on Android}
  \label{fig:google_voice search} %% label for entire figure
\end{figure}

GVS-Attack utilizes system build-in voice assistant functions. Voice assistant apps are handy little apps which can accept voice commands from the user and execute corresponding operations. In general, typical voice commands include dialing (like ``\emph{call Jack}''), querying (like ``\emph{what's weather tomorrow}'') and so on. Benefited from the development of speech recognition and other natural language processing techniques~\cite{clark2010handbook}, voice assistant apps facilitate users' daily operations and become an important selling point of smartphones. Every mainstream smartphone platform has its own built-in voice assistant app, such as Siri~\cite{siri_url} for iOS and Cortana~\cite{cortana_url} for Windows Phone. On Android platform, this function is implemented by Google Voice Search\footnote{Some people confuse Google Now with Google Voice Search, even the corresponding entries on Wikipedia. Actually Google Voice Search module can be invoked independently with disabling Google Now. See http://www.google.com/landing/now/} (see Figure \ref{fig:google_voice search}) which has been merged to Google Search app as a module from Android 4.1. Users can start Google Voice Search through touching the microphone icon of Google Search app widget, the shortcut named Voice Search, etc. Like other voice assistant apps, Google Voice Search can execute several kinds of operations without user touching the screen.

The basic idea behind GVS-Attack is to exploit the capabilities of Google Voice Search. Through Android Intent mechanism, VoicEmployer triggers Google Voice Search to the foreground, and then plays an audio file in the background. The audio file is special designed and the content is a voice command, such as ``\emph{call number 1234 5678}''. This command can be recognized by Google Voice Search and the corresponding operation would be executed. Through utilizing a vulnerability of status checking in Google Search app, zero permission based VoicEmployer can dial arbitrary malicious numbers even when the phone is securely locked. With ingenious designs, GVS-Attack can forge SMS/Email, access privacy information, transmit sensitive data and achieve remote control without any permission. Also context-aware information collection and analysis can assist GVS-Attack, which makes this attack more practical.

It is difficult to identify the malicious behaviors of VoicEmployer through current mainstream Android malware analysis techniques. From the aspect of permission checking~\cite{au2012pscout, felt2011android}, VoicEmployer doesn't need any permission. From the aspect of app dynamic behavior analysis~\cite{enck2010taintdroid, yan2012droidscope}, VoicEmployer doesn't execute malicious actions directly. The actual executor is Google Voice Search, a "trust" system built-in app module. In addition, played voice commands are outputted by the speaker and captured by the microphone as input. This inter-application communication channel and transmission media are beyond the control of Android OS. We tested several famous anti-virus apps (AVG, McAfee, etc.) to monitor the process of GVS-Attack. None of them can detect GVS-Attack and report VoicEmployer as a malware.

\vspace{5pt}
\textbf{Contributions.} We summarize this paper's contributions here:
\begin{itemize}
  \item \emph{New Attack Method and Surface.} To the best of our knowledge, GVS-Attack is the first attack method utilizing the speaker and voice assistant apps on mobile platforms. Also this attack can be launched by a totally zero permission Android malware and performs many malicious actions based on sensitive permissions bypassing, such as SMS/Email forging and privacy stealing.
%\vspace{-2pt}
  \item \emph{New Vulnerability.} We found a vulnerability of status checking in Google Search app (over 500 million installations), which can be utilized by GVS-Attack to dial arbitrary malicious numbers even when the phone is securely locked with password.
%\vspace{-2pt}
  \item \emph{Prototype Implementation and Evaluation in Real World.} We implemented a VoicEmployer prototype and carried out related experiments to demonstrate the feasibility of GVS-Attack. We also designed and tested related attack assisting schemes, including context-aware information analysis, sound volume setting, etc. These schemes make GVS-Attack more practical in real world.
\end{itemize}

\textbf{Roadmap.} The rest sections are organized as follows: Section \ref{sec:model} provides Google Voice Search related contents, including backgrounds, vulnerability analysis and the adversary model. The details of GVS-Attack are described in Section \ref{sec:attack}, which contains three different levels of attacks. In the following Section \ref{sec:evaluation}, a VoicEmployer prototype was implemented and related experiments were carried out. Section \ref{sec:defense} and Section \ref{sec:discussion} discuss corresponding defense strategies and some  related in-depth topics respectively. Previous research about sensor based attacks and Android app security analysis are reviewed in Section \ref{sec:relatedWork}. Section \ref{sec:conclusion} concludes this paper.

\section{Google Voice Search on Android}
\label{sec:model}
\subsection{Backgrounds}
\textbf{Google Services Framework.} Google Services Framework / Google Mobile Services are pre-installed on nearly all brands of Android devices. It can be treated as a suit of pre-installed apps developed by Google, including Google Play, Gmail, Google Search, etc.~\cite{gms_url} These killer apps (Google Search app has over 500 million installations just on Google Play) are so popular that even customized Android firmwares would keep them. Taking CyanogenMod\footnote{http://www.cyanogenmod.org/} as an example, due to licensing restrictions, Google Services Framework cannot come pre-installed with CyanogenMod, but these apps can still be installed via separate Google Apps recovery package~\cite{cm_google_url}.

\vspace{5pt}
\textbf{Android Intent Mechanism.} In Android OS, Intent mechanism allows an app start an activity / service in another app by describing a simple action, like "view map" and "take a picture"~\cite{AndroidIntent}. An Intent is a messaging object which declares a recipient (and contains data). VoicEmployer utilizes this mechanism to invoke Google Voice Search module of Google Search app. From Android OS' view, it only executes a normal operation to invoke a system built-in app module.

An intent filter is an expression in an app's manifest file that specifies the type of intents that the component would like to receive~\cite{AndroidIntent}. For example, in the manifest file of Google Search app, there defines hands free function (voice commands) related components can receive two kinds of actions, namely \texttt{ACTION\_VOICE\_\\COMMAND}~\cite{AndroidIntentList} and \texttt{ACTION\_VOICE\_SEARCH\_HANDS\_FREE}~\cite{AndroidRecognizerIntent}.

\subsection{Google Search App Vulnerability Analysis}
\label{subsec: vulnerability}

Google Voice Search is a voice assistant component / module of Google Search app. It is designed for hands free operations and can accept several kinds of voice commands, such as ``\emph{call Jack}'' and ``\emph{what's weather tomorrow}''. Google Voice Search runs in two modes: Voice Dialer -- Figure \ref{fig:dial_mode} and Velvet -- Figure \ref{fig:full_mode}. Voice Dialer mode only accepts voice dialing commands and Velvet mode is the full function mode. Google Voice Search can be invoked through Android Intent mechanism. First, a third-party app constructs an Intent based on \texttt{ACTION\_VOICE\_COMMAND} or \texttt{ACTION\_VOICE\_SEARCH\_HANDS\_FREE}, and then passes it to \texttt{startActivity()}. Android OS resolves this Intent and finds Google Search app can handle it. Then this Intent is passed to Google Search app. According to the current phone status\footnote{The code implementations of corresponding status checking are:  \texttt{KeyguardManager.isKeyguardLocked()}, \texttt{KeyguardManager.isKeyguardSecure()} and \texttt{PowerManager.isScreenOn()}.}, different modes will be started by Google Search app, namely:

\begin{enumerate}
  \item If the phone is unlocked and the screen is on, Velvet mode will be started.
      %\vspace{-2pt}
  \item If the phone is insecurely locked, namely sliding to unlock the screen without authentication, Voice Dialer mode will be started.
      %\vspace{-2pt}
  \item If the phone is securely locked, such as using password and pattern password, a voice warning will be played: "\emph{please unlock the device}".
\end{enumerate}

\begin{figure*}[htb]
\centering
  \includegraphics[width=15.8cm]{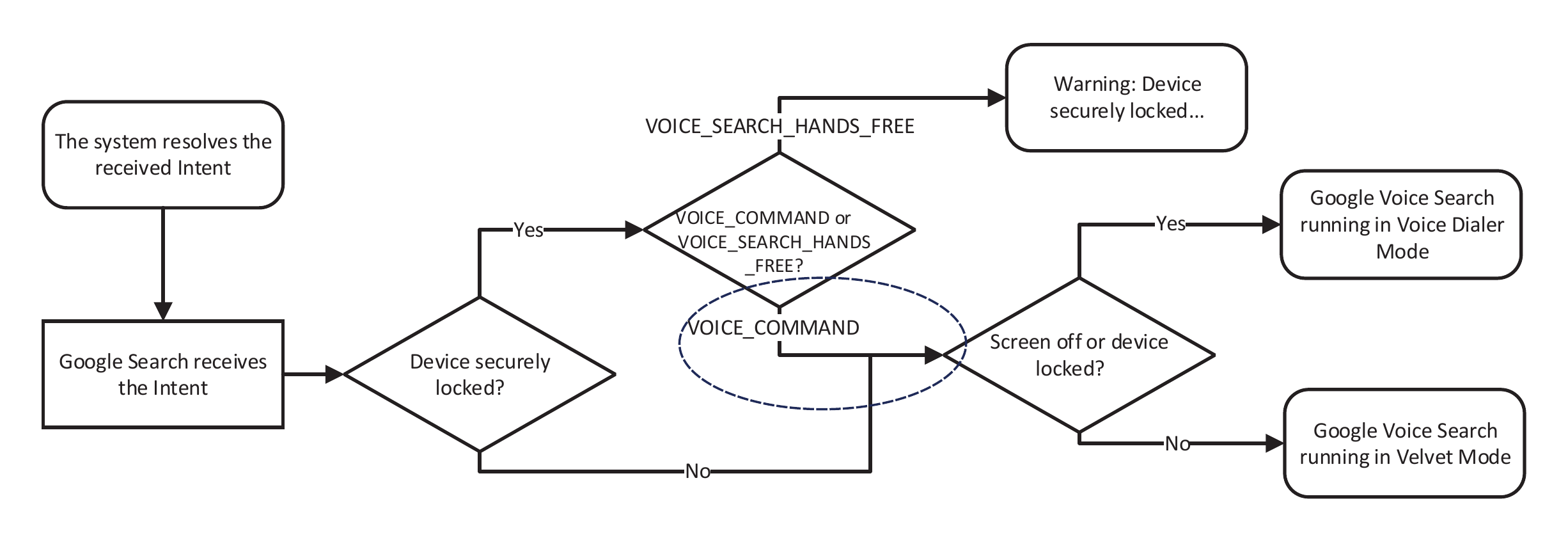}
  \vspace{-10pt}
  \caption{The Vulnerability of status checking in Google Search app}
  \label{fig:go_flow}
\end{figure*}

\textbf{Vulnerability in Google Search App.} Actually Voice Dialer mode can be started even the phone is securely locked. But this function is only designed for the Bluetooth headset hands free mode. This scene is reasonable, because the user must confirm the first connection requirement of the Bluetooth headset on his phone, which is a process of authorization. When the user long presses the button on his Bluetooth headset, if the phone is not in call process, the corresponding event can trigger passing an \texttt{ACTION\_VOICE\_COMMAND} based Intent to the OS. In Android source codes, \texttt{HeadsetStateMachine.java} defines and handles related operations~\cite{HeadsetStateMachine}.

But we found a vulnerability that Google Search app doesn't strictly check whether the phone is connected with a Bluetooth headset. This vulnerability results in that a third-party app can pass an \texttt{ACTION\_VOICE\_COMMAND} based Intent to the OS and triggers Voice Dialer mode of Google Voice Search, even when the phone is securely locked. Through decompiling the apk file of Google Search app (version 3.3.11.1069658.arm\footnote{Following versions have been applied code obfuscation techniques, and this vulnerability still exists.}, released on March 14th 2014), we analyzed related code workflows and located this vulnerability. Figure \ref{fig:go_flow} shows the simplified logical flow of Google Search app handling voice assistant type of Intent. The area bounded by the dotted line is the location of the vulnerability of status checking.

\subsection{Adversary Model}
GVS-Attack needs to be launched by an Android malware (called VoicEmployer) which has been installed on the user's phone. This malware could disguise as a normal app and contains attack modules. Since VoicEmployer doesn't need any permission, so this disguise should be quite easy. After VoicEmployer is opened by the user (namely the victim) once, the subsequent attack processes don't need to interact with the victim. The scene of attack scenario is when the victim is not using his Android phone.

\vspace{5pt}
\textbf{Assumptions.} GVS-Attack contains three different levels: Basic Attack, Extended Attack and Remote Voice Control Attack.
\begin{enumerate}
  \item The victim's Android phone contains complete Google Services Framework. The reason is VoicEmployer needs to invoke Google Voice Search (and Google speech recognition / synthesis service) to execute attacks.
      %\vspace{-2pt}
  \item The victim's Android phone is securely locked or not. The secure lock status corresponds to Basic Attack and the insecure lock status corresponds to all levels of attacks.
\end{enumerate}

The requirement of the first assumption is quite weak, because most smartphone vendors pre-install Google Services Framework on their products. To the second assumption, through utilizing the vulnerability of Google Search app, Basic Attack of malicious number dialing can be launched even when the phone is securely locked. To launch Extended Attack and Remote Voice Control Attack, an additional assumption is that the victim doesn't use secure screen lock functions, such as password and pattern password. Because Google Voice Search cannot run in Velvet mode without unlocking the screen. Actually for convenience, many users (more than 30\%) would not like to use secure screen lockers~\cite{smartphone_sec_survy,mobile_privacy_tip}.

\vspace{5pt}
\textbf{Zero Permission.} In our designs, GVS-Attack can be launched by VoicEmployer \textbf{without any permission}. Android OS uses a permission mechanism to enforce restrictions on the specific operations that a particular process can perform. The permission abusing problem has been noticed long before, that some sensitive permissions (such as \texttt{CAMERA}, \texttt{RECORD\_AUDIO} and \texttt{READ\_CONTACTS}) are utilized by malicious apps to collect user privacy and achieve other illegal targets. Because this problem is so widespread, some users will pay attention to an unknown or new app requiring sensitive permissions~\cite{felt2012android}.

Instead of requesting these permissions explicitly, our attack can perform sensitive actions protected by permissions only using the speaker. According to the current Android permission mechanism, playing audio doesn't require any permission.

\vspace{5pt}
\textbf{Context-aware Information Collection and Analysis.} Since played voice commands may be heard and interrupted by the victim, VoicEmployer needs to analyze the current environment to decide whether launching attacks, that is collecting information through sensors and legal Android SDK API implementations. On mobile platforms, the feasibility of context-aware information analysis has been demonstrated in several previous research~\cite{gellersen2002multi, lee2011activity, schlegel2011soundcomber}. Actually this target also could be achieved \textbf{without any permission}. For example, light sensor and accelerometer can be used to analyze the external environment. And the internal status of the phone can be checked through analyzing CPU workload, memory workload, etc., all of which can be accessed free from any permission.

\vspace{5pt}
\textbf{Typical Attack Scene.} Due to job requirements or just personal habits, many people don't turn off their phones even they are sleeping~\cite{smith2011americans}. In the early morning (before dawn, such as 3 AM), the victim is likely to be in deep sleep and the environment is quiet. The sound volume of played voice commands could be very low and still be recognized by Google Voice Search. According to sleep related medical research, nocturnal awakenings usually occur with noise levels greater than 55 $dB$~\cite{muzet2007environmental}. It means voice commands of low sound volume will not awaken the victim. This situation provides an ideal scenario for GVS-Attack launching. The following sections are based on this typical scene.

%but actually it is not so secure as user general option. Our attacks invoke Google Voice Search to do several malicious behaviors.

% vim: tw=0

\section{Attacks}
\label{sec:attack}
%\todo{I think we can organize these three attacks based on different assumptions. For example, starting from the zero-permission, then add extra one that can lead a stronger attack, and so on.}

The basic idea behind GVS-Attack is to exploit the capability of Google Voice Search and bypass Android permission checking mechanism. VoicEmployer can trigger Google Voice Search to the foreground and then plays an audio file in the background. This audio file records a voice command, like "\emph{call number 1234 5678}". Android system allows an app playing audio files in a service, which means it doesn't need to start an activity and show UI. This voice command can be recognized by Google Voice Search and the corresponding operation will be executed. Because Google Voice Search can accept several kinds of commands including sending SMS, opening maps, making notes, etc., attacks can be varied. There are two highlights in our attacks:

\emph{Attack in Any Situation.} Through utilizing the vulnerability of status checking in Google Search app, zero permission based VoicEmployer can dial arbitrary malicious numbers even when the phone is securely locked.

\emph{Remote Data Transmission and Voice Control.} User Data can be transmitted via the call channel. Also the attacker can control the victim's Android phone remotely. Generally a mobile malware needs quite sensitive privileges (even root) to achieve remote control. However our attack doesn't need any permission.

\begin{figure}[htb]
  \includegraphics[width=8.5cm]{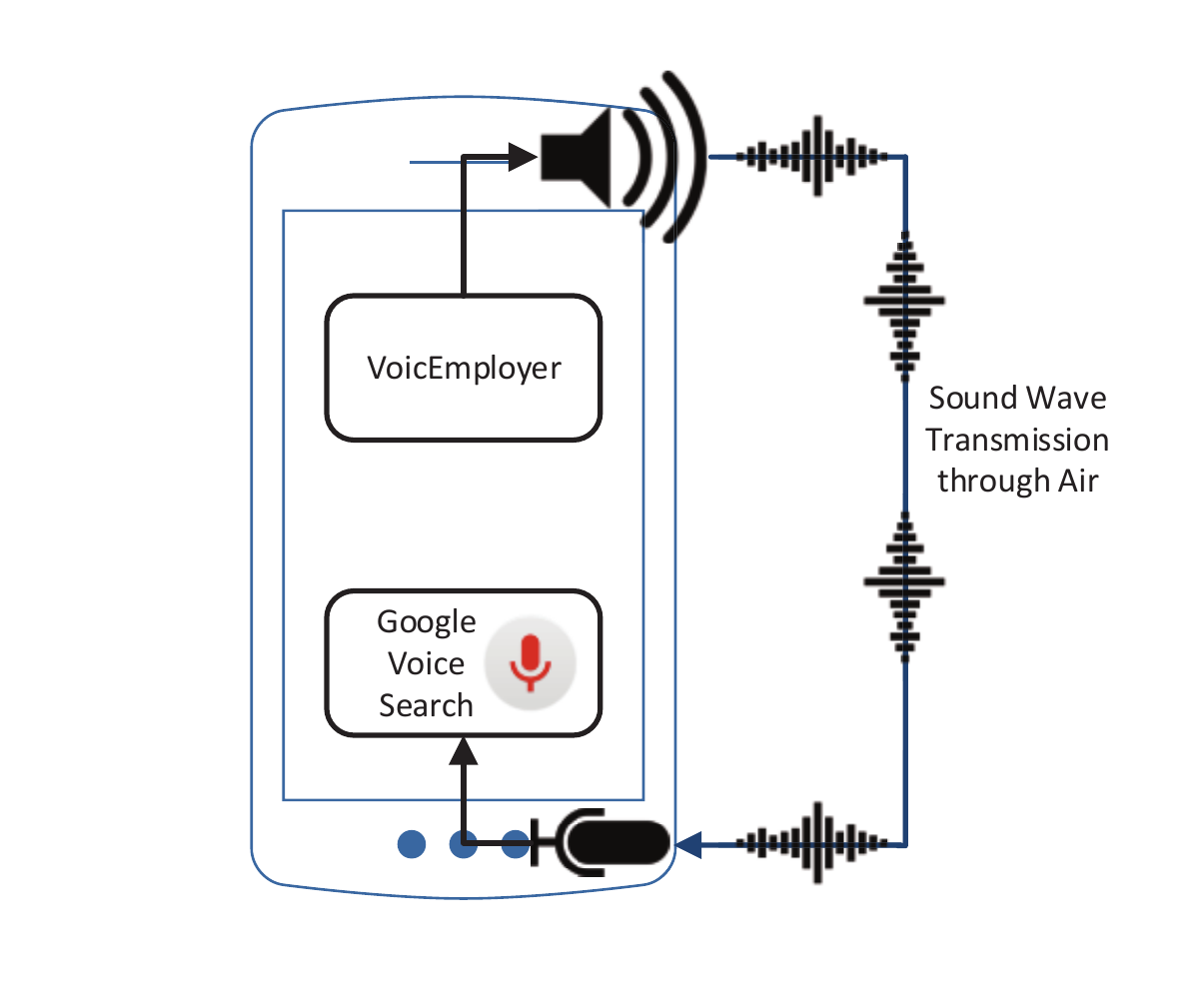}
  \vspace{-25pt}
  \caption{Inter-Application Communication Channel of GVS-Attack}
  \label{fig:communication}
\end{figure}

\vspace{5pt}
%Also if VoicEmployer is assigned more permissions, with special designs, GVS-Attack will become more harmful.
GVS-Attack provides a \textbf{new inter-application communication channel} for mobile malware attacks. In the process of GVS-Attack, the input of microphone (Google Voice Search) comes from the output of speaker (VoicEmployer). This information transmission process is beyond the control of Android OS, namely an uncontrolled physical communication channel or a kind of covert channel. Figure \ref{fig:communication} shows this communication channel. Android OS cannot distinguish the source of voice commands, namely whether they come from the user or an internal app. This new attack channel is totally different from previous research on security of application communications~\cite{kantola2012reducing, chin2011analyzing}.
%Similar potential covert channels should be noticed.

%\vspace{5pt}
In the following subsections, three types of attack designs are described in details. Basic Attack can be launched by VoicEmployer even when the phone is securely locked. With the assumption that the victim doesn't use secure screen lock functions, GVS-Attack are extended to bypass several sensitive permissions in Extended Attack. Combined previous two attacks, the attacker can interact with the phone, which results in Remote Voice Control Attack.

\subsection{Basic Attack: Malicious Number Dialing}
\label{sec:attack1}
Basic Attack achieves arbitrary malicious number dialing. That is to say the \texttt{CALL\_PHONE} permission is bypassed directly. The vulnerability of Google Search app makes this attack still valid when the phone is securely locked. Without dismissing screen keyguard, Google Voice Search would run in Voice Dialer mode which only accepts dialing type of commands, like Figure \ref{fig:dial_mode} shows. Before launching attacks, we answer two questions first: how to invoke Google Voice Search and how to prepare voice command files.

\textbf{Google Voice Search Invoking.} When both the external environment and the phone internal status reach the predefined trigger threshold, VoicEmployer will invoke Google Voice Search through Android Intent mechanism. Actually VoicEmployer just needs to construct an Intent based on \texttt{ACTION\_VOICE\_COMMAND} or \texttt{ACTION\_VOICE\_SEARCH\_HANDS\_FREE}, and passes it to API \texttt{startActivity()}. When the phone is securely locked, Google Voice Search is designed keeping the disable status and requiring the user to dismiss keyguard. But if the Intent is constructed using \texttt{ACTION\_VOICE\_COMMAND}, recalling Section~\ref{subsec: vulnerability}, this restriction can be bypassed.

\textbf{Voice Command Files Preparation.} After Google Voice Search start, VoicEmployer plays an audio file (voice commands) as a background service. These played voice commands could be packaged in VoicEmployer as a part of apk resource, but this method would increase the size of installation package. A more covert solution is utilizing Google Text-to-Speech (TTS)~\cite{AndroidTextToSpeech} service which is a system built-in service used for translating text contents to speech. TTS service is based on Internet, but it doesn't require that the app invoking TTS declares the \texttt{INTERNET} permission. Therefore VoicEmployer can execute speech synthesis and save results as an audio file in its own data folder (internal storage) dynamically.

\vspace{5pt}
\emph{Attack Launching.} The scene of attack scenario is when the victim is not using his Android phone, namely the phone is in system sleep status (screen off). In previous preparations, VoicEmployer synthesizes two audio records like ``\emph{call 1234 5678}'' and ``\emph{OK}'', in which 1234 5678 could be a malicious number.

When this attack is launched, Voice Dialer mode of Google Voice Search is started first and VoicEmployer plays the audio record -- ``\emph{call 1234 5678}''. Then Voice Dialer recognizes this command and gives a voice feedback ``\emph{do you want to call 1234 5678, say OK or cancel}''.  VoicEmployer plays another audio record ``\emph{OK}''. After receiving this confirmation, finally Voice Dialer makes a call to 1234 5678. Since 1234 5678 is a malicious number, so if the dial is connected, the victim's phone account will be charged with premium-rate service fee. Therefore even only dialing commands can be accepted, the victim has to afford to lose money and his phone number is leaked.

\subsection{Extended Attack: Sensitive Permissions Bypassing}
\label{sec:attack2}
Extended Attack needs to invoke Velvet mode of Google Voice Search, like Figure \ref{fig:full_mode} shows. Therefore this attack occurs in the situation that the victim doesn't use secure screen lock functions on his Android phone. In addition, VoicEmployer needs to unlock and turn on the screen itself.

In general, an app needs the \texttt{WAKE\_LOCK} permission to wake the phone from sleep status, and then sets flag \texttt{LayoutParams}.\texttt{FL\\AG\_DISMISS\_KEYGUARD} to dismiss the screen keyguard. Since \texttt{WAKE\_LOCK} is not a sensitive permission and doesn't correlate with user data, so this requirement is not exorbitant. Actually there is a tricky implementation to bypass this permission. That is, VoicEmployer invokes Google Voice Search once and does nothing until it exits for timeout. At this moment, the phone has been in waked status (the screen is on) and the keyguard has be dismissed. It can be treated as that VoicEmployer exploits the \texttt{WAKE\_LOCK} permission of Google Voice Search to wake the phone.

\vspace{5pt}
\textbf{User Sensitive Data Collection.} Running in Velvet mode, Google Voice Search can accept more types of voice commands~\cite{voiceAndroid_url}, not just dialing commands. Different voice commands can lead to different information leakages. Some commands make that specific operations are executed, such as sending SMS. Other querying type of commands will trigger voice feedbacks from Google Voice Search, such as ``\emph{what is the time?}'' will get ``\emph{the time is 9:39 pm}''. Typical information leakages include (\emph{italic sentences} are voice commands or voice feedback):

\begin{itemize}

\item ``\emph{Email to [contacts], subject ``meeting cancel'', message ``tomorrow's meeting has been canceled''.}'' This command results in sending an Email to the contacts with the above subject and message, which means it can be used to forge Emails with any contents.
%\vspace{-2pt}
  \item ``\emph{Send SMS to number 1234 5678 ``confirm subscribe to weather forecast service''.}'' This command results in sending an SMS to number 1234 5678 (prepared by the attacker), which means it can be used to forge SMS and subscribe to premium-rate services.
%\vspace{-2pt}
  \item ``\emph{What is my next meeting?}'' $\Rightarrow$``\emph{Your next calendar entry is tomorrow 10 AM. The tile is ``Meet with boss''.}'' The victim's calendar schedule is leaked through voice feedback.

%\vspace{-2pt}
  \item ``\emph{What is my IP address?}'' $\Rightarrow$ ``\emph{Your public IP address is 111.222.111.222.}'' The victim's IP address is leaked through voice feedback.
%\vspace{-2pt}
  \item ``\emph{Where is my location?}'' $\Rightarrow$ ``\emph{Here is a map of Brooklyn District.}'' The victim's location (district level) is leaked through voice feedback.
\end{itemize}

Actually it equals that VoicEmployer exploits the capability of Google Voice Search and bypasses Android permission checking mechanism. From the Android OS's view, VoicEmployer just passed an Intent and plays some audio files. Table \ref{tab:permission_bypass} shows the permissions which can be bypassed through GVS-Attack. Some voice commands are not listed, because they need interaction operations (touch the screen) of the user, such as ``\emph{create a calendar event ...}'' and ``\emph{post to Google plus ...}''.

\begin{table}[htb]
  \caption{Permissions Bypassed by GVS-Attack}
  \vspace{5pt}
  \centering
  \begin{tabular}{|p{3.5cm}|p{4.2cm}|}
   \hline
    \multicolumn{1}{|c|}{\textbf{Voice Command}} & \multicolumn{1}{c|}{\textbf{Bypassed Permission(s)}}\\
%    \hline
%    Call [number] & \texttt{CALL\_PHONE} \\
    \hline
    Call ... & \texttt{READ\_CONTACTS}, \texttt{CALL\_PHONE}\\
    \hline
    Listen to voicemail & \texttt{WRITE\_SETTINGS}, \texttt{CALL\_PHONE}\\
    %\hline
    %Post to Google+ ... & \texttt{GET\_ACCOUNTS}, \texttt{INTERNET}\\
    \hline
    Browse to Google dot com & \texttt{INTERNET}\\
    \hline
    Email to ... & \texttt{READ\_CONTACTS}, \texttt{GET\_ACCOUNTS}, \texttt{INTERNET}\\
%    \hline
%    Send SMS to [number] ... & \texttt{SEND\_SMS}\\
    \hline
    Send SMS to ...& \texttt{READ\_CONTACTS}, \texttt{WRITE\_SMS}, \texttt{SEND\_SMS}\\
    \hline
    Set alarm for ... & \texttt{SET\_ALARM} \\
    \hline
    Note to myself ... & \texttt{GET\_ACCOUNTS}, \texttt{RECORD\_AUDIO}, \texttt{INTERNET}\\
    \hline
    What is my next meeting? & \texttt{READ\_CALENDAR}\\
     \hline
    Show me pictures of ... &  \texttt{INTERNET} \\
    \hline
    What is my IP address? & \texttt{ACCESS\_WIFI\_STATE}, \texttt{INTERNET}\\
    \hline
    Where is my location?  & \texttt{ACCESS\_COARSE\_LOCATION}, \texttt{INTERNET}\\
    \hline
    How far from here to ...? & \texttt{ACCESS\_FINE\_LOCATION}, \texttt{INTERNET}\\
   \hline
\end{tabular}
\label{tab:permission_bypass}
\end{table}

% \texttt{BIND\_NOTIFICATION\_LISTENER\_SERVICE}
\vspace{5pt}
\textbf{Remote Data Transmission.} GVS-Attack can dial a malicious number through playing ``\emph{call ...}'', when this call is answered by an auto audio record machine, actually the data transmission channel has been built. Any audio type of data can be transferred through this channel instead of commonly used Internet connection.

\emph{Google Voice Search running during calling period.} In Android OS, the audio recording method is synchronized, which means multiple apps cannot access the microphone at the same time on SDK API level. But as an essential system module, the phone call function of Android is based on hardware-level implementations instead of invoking \texttt{MediaRecorder} / \texttt{AudioRecord} class. Android OS only sends control signals and related hardware (GSM module, microphone, speaker, etc.) complete audio data processing functions directly~\cite{Telephony}. Therefore during the phone call period, another app can also access the microphone through Android SDK API. It means Google Voice Search still can be triggered to the foreground to accept voice commands\footnote{Google Voice Search is an Internet based service. With 3G/4G or Wi-Fi data connection, the phone can make phone call and access the Internet at the same time. To 2G data connection (GPRS and EDGE), the two functions are conflicting.} normally.

Therefore, to querying commands, the feedback voice can be got by the attacker through this call channel. Based on speech recognition techniques, the attacker can get the corresponding text information directly. More details, VoicEmployer invokes Google Voice Search and then plays voice command ``\emph{call number 1234 5678}'', in which 1234 5678 is the number of a malicious auto audio record machine. After this call connects, VoicEmployer invokes Google Voice Search and plays querying commands again, like ``\emph{where is my location}''. The feedback voice -- ``\emph{here is a map of Sha Tin district}'' of Google Voice Search is transmitted to the attacker through the call channel.

If VoicEmployer could get more permissions, more sensitive information of the victim would be leaked. A natural idea is to play audio records on external storage directly. These audio records are recorded by the victim previously and probably contain sensitive information, like a business negotiation. This operation is based on the \texttt{READ\_EXTERNAL\_STORAGE}\footnote{The \texttt{READ\_EXTERNAL\_STORAGE} permission is added from Android 4.1. But before Android 4.4, users must turn on this permission checking option manually. Otherwise, an app still can read all data on external storage without any permission.} or \texttt{WRITE\_EXTERNAL\_STORAGE} permission. Both of them allow an app read all data on external storage, not just its own folder. For text files, they can be translated to audio data through TTS service. For example, with the \texttt{READ\_CONTACTS} permission, VoicEmployer can get phone contacts of the victim. Then it utilizes TTS service to speak these texts of contacts directly. So the attacker can get a complete copy of the victim's contacts through the call channel. Similar potential risks include \texttt{READ\_CALL\_LOG}, \texttt{READ\_SMS}, etc.

The voice channel is not just used for transferring text information. Actually any type of files can be translated to hex coding formats, hence in theory any file could be transmitted in the form of audio coding (or even read hex codes directly). So when the attacker receives all hex codes of a photo on external storage through the call channel, he can restore it easily. If this photo contains sensitive information, such as selfie, leaked information will be quite harmful to the victim. Also in practical attacks, compressed encoding and checking mechanism need to be considered.

\subsection{Remote Voice Control Attack: Real World Case Study}
\label{sec:attack3}
Remote Voice Control Attack combines previous Basic Attack and Extended Attack. This attack also needs to invoke Velvet mode of Google Voice Search.

In Remote Voice Control Attack, after VoicEmployer triggers a malicious number dialing, the attacker answers this call directly. During the calling period, VoicEmployer invokes Google Voice Search periodically. When the attacker speak some voice commands, these commands will be played by the speaker (headset) of victim's phone. Then Google Voice Search recognizes these commands and executes corresponding operations. It means that the attacker can interact with the victim's phone through Google Voice Search, namely remote voice control.

Previous research~\cite{zhou2012dissecting} showed more than 90\% of Android malwares would turn the compromised phones into a botnet controlled through network or short messages. These controls were based on corresponding communication permissions (such as \texttt{INTERNET} and \texttt{SEND\_SMS}) or even root privilege. But our remote control method exploits the capability of Google Voice Search to build the communication channel and touch sensitive data. These targets are completed without corresponding permissions.

With special designs, Remote Voice Control Attack could become quite powerful. One example of voice interaction is the command ``\emph{How far from here to Lincoln Memorial by car?}''. Google Voice Search provides a voice feedback like ``\emph{The drive from your location to Lincoln Memorial is 17.6 kilometers}''. Then the attacker can use the successive approximation method to ask similar questions of different locations and until get a feedback like ``\emph{The drive from your location to White House is 120 meters}''. At this moment, the attacker gets the accurate location of the victim, which is quite dangerous. Based similar methods, more dangerous attacks could be designed.

Another example is that the attacker can leave a note to the victim using the command ``\emph{Note to self: You have been hacked}''. Google Voice Search would make a note (send an Email to the victim himself) with the content ``\emph{You have been hacked}''. In the morning, after the victim get up, he will find his phone appears such a strange note. Also based on this thought, some complex social engineering attacks~\cite{hadnagy2010social} could be designed and launched.

% vim: tw=0

\section{Implementation and Analysis}
\label{sec:evaluation}

\subsection{Overall Structure and Attack Experiments}
\label{subsec:structure}
Based on previous descriptions, we implemented a VoicEmployer prototype to demonstrate our attack schemes. Our implementation of VoicEmployer contains 5 modules: MainActivity, AlarmReceivor, EnvironmentService, WakedActivity and VoiceCommandService.

\begin{itemize}
  \item \textbf{\texttt{MainActivity}} shows the normal starting UI which seems like a normal app. Its main functions include registering an alarm and preparing audio files of voice commands.
%\vspace{-2pt}
  \item \textbf{\texttt{AlarmReceivor}} is designed to receive alarm events and start EnvironmentService.
%\vspace{-2pt}
  \item \textbf{\texttt{EnvironmentService}} is designed for context-aware information collection and analysis. Analyzed aspects include light sensor, accelerometer, \texttt{/proc/}, system workloads, etc. All of these collection operations don't require any permission. If the analysis results reach a predefined threshold, WakedActivity will be started. Related implementation details are described in Section \ref{subsec:content_analysis}.
%\vspace{-2pt}
  \item \textbf{\texttt{WakedActivity}} is designed to dismiss insecure screen keyguard through setting \texttt{LayoutParams}.\texttt{FLAG\_DISMISS\\\_KEYGUARD}. This function only can be implemented in Activity components. Without dismissing screen keyguard, Velvet mode cannot be started. \emph{Note: This module is not necessary for Basic Attack.}
%\vspace{-2pt}
  \item \textbf{\texttt{VoiceCommandService}} is designed to invoke Google Voice Search and play voice command files. Voice command playing will be delayed a short while to wait Google Voice Search ready. To Remote Voice Control Attack, Google Voice Search will be invoked periodically.
\end{itemize}

%\begin{description}
%  \item[\texttt{MainActivity}] MainActivity shows the normal starting UI which seems a normal app. Its main functions include registering an alarm and preparing audio files of voice commands.
%  \item[\texttt{AlarmReceivor}] AlarmReceivor is designed to receive alarm events and start EnvironmentService.
%  \item[\texttt{EnvironmentService}] EnvironmentService is designed for context-aware information collection and analysis. Analyzed aspects include light sensor, accelerometer, \texttt{/proc/}, system workloads, etc. All of these collection operations don't need to use any permission. If the analysis results reach a predefined threshold, WakedActivity will be started. Related implementation details are described in Section \ref{subsec:content_analysis}.
%  \item[\texttt{WakedActivity}] WakedActivity is designed to dismiss insecure screen keyguard through setting \texttt{LayoutParams}.\texttt{FLAG\_\\DISMISS\_KEYGUARD}. This function only can be implemented in Activity components. Without dismissing screen keyguard, Velvet mode cannot be started. \emph{Note: This module is not necessary for Basic Attack.}
%  \item[\texttt{VoiceCommandService}] VoiceCommandService is designed to start Google Voice Search and play voice command files. Voice command playing will be delayed a short while to wait Google Voice Search ready. To Remote Voice Control Attack, Google Voice Search will be started periodically.
%\end{description}

\begin{figure}[htb]
\centering
  \includegraphics[width=8.1cm]{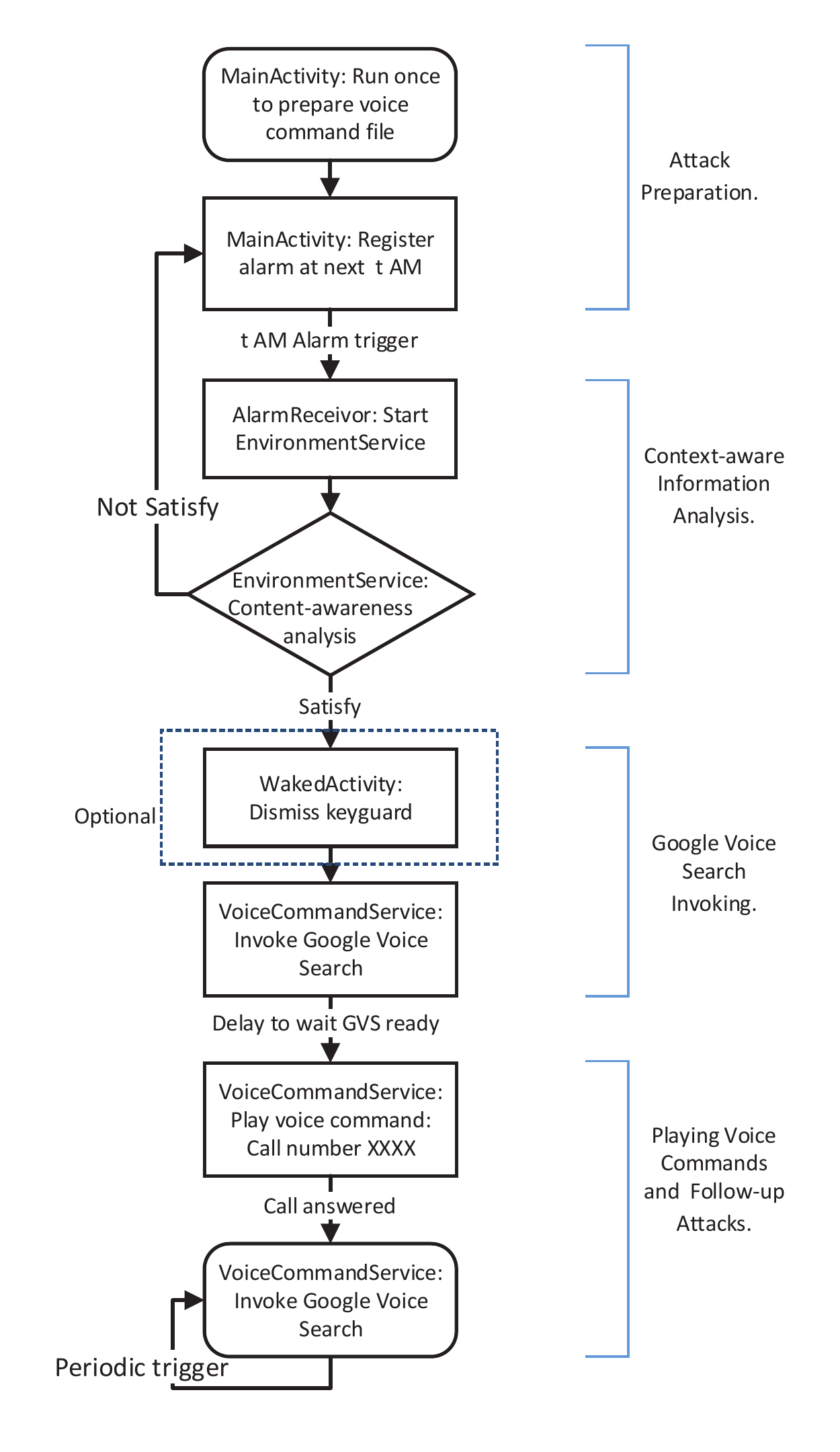}
  \vspace{-10pt}
  \caption{Workflow of VoicEmployer Implementation}
  \label{fig:implementation}
\end{figure}

Taking Remote Voice Control Attack as an example, Figure \ref{fig:implementation} shows the workflow of VoicEmployer implementation. It utilizes the alarm mechanism to trigger attack related modules to avoid constant background services.

%To make GVS-Attack work well, the victim's phone should use Android 4.1 or upper versions. Because Google Search app running on Android 4.0 or lower versions cannot be updated to the newest version. It is a measure of encouraging system update to avoid version fragmentation. The experiment version of Google Search app was 3.3.11.1069658.arm (released on March 14th 2014). We tested VoicEmployer prototype on Samsung Galaxy S3 GT-I9300 (CyanogenMod 4.4.2), Meizu MX2 (Meizu official Android 4.2.1) and Motorola A953 (CyanogenMod 10, Android 4.1.1). GVS-Attack can be launched on all of them successfully. On Samsung Galaxy S3, when VoicEmployer used \texttt{ACTION\_VOICE\_SEARCH\\\_HANDS\_FREE} to construct the Intent, S Voice app\footnote{http://www.samsung.com/global/galaxys3/svoice.html} (a built-in voice assistant developed by Samsung) would be started, not generally Google Voice Search, but attack processes are the same. This change comes from vendor customizations, that is the Intent recipient of \texttt{ACTION\_VOICE\_SEARCH\_HANDS\_FREE} is modified. Also \texttt{ACTION\_VOICE\_COMMAND} based Intent works well and Google Voice Search can be started normally. Table \ref{tab:GVS_result} summarizes the experiment results.

\vspace{5pt}
\textbf{Attack Results.} In order to make GVS-Attack work well, the victim's phone should use Android 4.1 or upper versions. Because Google Search app running on Android 4.0 or lower versions cannot be updated to the newest version which supports more voice commands. It is a measure of encouraging system update to avoid version fragmentation. The experiment versions of Google Search app were 3.3.11.1069658.arm and 3.4.16.1149292.arm, which were released on March 14th 2014 and May 6th 2014 respectively. We tested VoicEmployer prototype on Samsung Galaxy S3 GT-I9300, Meizu MX2 and Motorola A953. GVS-Attack can be launched on all of them successfully. On Samsung Galaxy S3 using the official firmware, S Voice app~\cite{svoice_url} (a built-in voice assistant developed by Samsung) would be started, not generally Google Voice Search, but attack processes are similar. This change derives from that Samsung sets a higher priority in the manifest file of S Voice app to respond \texttt{ACTION\_VOICE\_COMMAND} and \texttt{ACTION\_VOICE\_SEARCH\_HANDS\_FREE}. Table \ref{tab:GVS_result} summarizes the experiment results.

\begin{table}[htb]
  \caption{GVS-Attack Experiments}
  \vspace{5pt}
  \centering
  \begin{tabular}{|l|l|c|}
   \hline
     \textbf{Phone Model} & \textbf{Android Version} & \textbf{Attack Result}\\
  \hline
    \multirow{2}{*}{Samsung Galaxy S3}	& CyanogenMod 4.4.2 & success\\
  \cline{2-3}
  	& Samsung Official 4.3 & success\\
  \hline
  Meizu MX2 &Meizu official 4.2.1 & success\\
  \hline
  Motolora A953&CyanogenMod 4.1.1& success\\
 \hline
\end{tabular}
\label{tab:GVS_result}
\end{table}

\vspace{5pt}
\textbf{Speech Recognition Accuracy.} We consider Command Success Rate (CSR) instead of Word Error Rate (WER)~\cite{schalkwyk2010your}, because executing results are what we only care. Actually test results showed CSR was nearly 100\%. This high percentage comes from three reasons: 1. the attack scene is the quiet environment, which means there is no interfering noise; 2. these command sentences are simple English, which don't lead to ambiguity; 3. VoicEmployer utilizes Google TTS to synthesize the command speech, which removes accent and pronunciation problems.

\vspace{5pt}
\textbf{Detection by Anti-Virus Apps.} We tested the following anti-virus apps for Android platform: AVG AntiVirus\footnote{https://www.avgmobilation.com/}, McAfee Antivirus \& Security\footnote{https://www.mcafeemobilesecurity.com/}, Avira Antivirus Security\footnote{http://www.avira.com/en/free-antivirus-android}, ESET Mobile Security \& Antivirus\footnote{http://www.eset.com/us/home/products/mobile-security-android/} and Norton Mobile Security\footnote{https://mobilesecurity.norton.com/}. After executing threat scanning, none of them reported VoicEmployer prototype as a malware. Also in the attack processes, none of their real-time monitoring modules can detect GVS-Attack's malicious behaviors.

\subsection{Context-aware Information Collection and Analysis Experiments}
\label{subsec:content_analysis}
The scene of GVS-Attack scenario is when the victim is not using his Android phone. Especially in the early morning (before dawn, such as 3 AM), the victim is likely to be in deep sleep. To avoid that played voice commands may be found and interrupted by the victim, VoicEmployer needs to analyze the current environment before deciding whether launching attacks, including external environment and phone status. This target can be achieved through sensors and legal Android SDK API implementations without any permission. If the analysis result shows the victim is using his phone, VoicEmployer will keep inactive status. Otherwise, GVS-Attack could be launched. Analysis aspects and criteria include:

\begin{itemize}
  \item Light sensor. Get the current external environment brightness: if the brightness is very low (such as less than 5\% of max), it means the phone is probably in pockets / bags or the victim has turned off the room lamp at night.
%\vspace{-2pt}
  \item Accelerometer. Get the current posture information of the phone: if the result shows the phone is in static status (namely stable readings), it means the victim is probably not holding / taking the phone.
%\vspace{-2pt}
  \item Android SDK class - \texttt{PowerManager}. Check the screen is on or off: if the screen is off, it means the victim is probably not using the phone.
%\vspace{-2pt}
  \item Android SDK class - \texttt{SimpleDateFormat}. Get the current system time: for example, 3 AM means the victim is likely to be in deep sleep.
%\vspace{-2pt}
  \item Read \texttt{/proc/stat} and \texttt{top} command. Analyze the current CPU workload: if the workload is low (such as less than 50\%), it means the victim is probably not using his phone or running large apps. \emph{Note: This value is affected by hardware configurations and background services.}
%\vspace{-2pt}
  \item Read \texttt{/proc/meminfo} and \texttt{top} command. Analyze the current RAM usage: if the usage is low (such as less that 50\%), it means the victim is probably not using his phone or running large apps. \emph{Note: This value is affected by hardware configurations and background services.}
\end{itemize}

Based on above analyses, we designed and carried out context-aware information analysis experiments using EnvironmentService module in 3 typical scenes:

\begin{enumerate}
  \item The user is using his phone to play games (take Angry Birds Rio\footnote{http://www.rovio.com/en/our-work/games} for example) with the room lamp is on.
  %\vspace{-2pt}
  \item The user is walking on the street, while the phone (screen-off) is in his trouser pocket.
  %\vspace{-2pt}
  \item $[$\emph{Target Scene}$]$ The phone (screen-off) is put on a horizontal table with the room lamp is off.
\end{enumerate}

The experiment was based on Meizu MX2 which has current mainstream hardware configurations (ARM Cortex-A9 quad-core 1.6 GHz, 2 GB LPDDR2 RAM). Before every scene's test, the phone was restarted for cutting out distractions of irrelevant factors. Table \ref{tab:content_aware} shows test results in 30 seconds. We can find that some item values of Scene 3 (target scene) are significantly different from those ones of Scene 1 and Scene 2. It demonstrates that context-aware information analysis can be used to detect the current environment and assist GVS-Attack practically.

\begin{table}[htb]
\caption{Content-aware Information Analysis}
  \vspace{5pt}
  \centering
  \begin{tabular}{|c|c|c|c|c|}
  \hline
    \multicolumn{2}{|c|}{\textbf{Method / Tool}} &\textbf{ Scene 1} & \textbf{Scene 2} & \textbf{Scene3}\\
  \hline
  \multirow{3}{*}{Light Sensor}	& Max&	304 &1 & 0\\
  %\cline{2-5}
  	&Min & 126 & 0& 0\\
  %\cline{2-5}
   &Avg & 227.95 & 0.5& 0\\
  \hline
  \multirow{9}{*}{Accelerometer}	& X-axis Max	& 7.78 & 18.19 & 0.15 \\
  %\cline{2-5}
    & X-axis Min & 5.62 &-0.06 & 0.01\\
   %\cline{2-5}
    & X-axis Avg &6.54 & 8.61 & 0.10\\
   \cline{2-5}
    & Y-axis Max & -0.50 & 14.56 & 0.11 \\
   %\cline{2-5}
    & Y-axis Min & -1.97 & -2.15 & -0.10\\
   %\cline{2-5}
    & Y-axis Avg & -1.10& 4.78 & 0.01\\
   \cline{2-5}
   & Z-axis Max &9.46 & 6.70 & 10.34 \\
   %\cline{2-5}
   & Z-axis Min & 5.88 & -12.00 & 10.04\\
   %\cline{2-5}
   & Z-axis Avg & 7.60 & -0.18 & 10.22\\
   \hline
    \multirow{3}{*}{CPU Workload}	& Max&	59\% & 43\%& 40\%\\
  %\cline{2-5}
  	&Min & 32\% & 22\% & 18\%\\
  %\cline{2-5}
   &Avg & 43\% & 29\% & 28\%\\
   \hline
    \multirow{3}{*}{Memory Usage}	& Max&	66.2\% & 53.4\% & 52.3\% \\
  %\cline{2-5}
  	&Min &  66.0\%& 53.2\% & 52.0\%\\
  %\cline{2-5}
   &Avg & 66.0\%& 53.3\% & 52.0\%\\
  \hline
  \multicolumn{2}{|c|}{ Screen Status} & on & off & off\\
  \hline
\end{tabular}
\label{tab:content_aware}
\end{table}

%  \tabincell{c}{Microphone \\(not compulsory)} & \texttt{RECORD\_AUDIO} & \tabincell{l}{Analyze the environment sound decibel: confirm the environment is very quiet, which \\means there should not exist human activities.}\\
%\begin{table*}[htb]
%  \caption{Context-aware Information Collection and Analysis without Permission}
%  \vspace{5pt}
%  \centering
%  \begin{tabular}{|l|l|}
%   \hline
%  \textbf{Sensors/Methods} & \textbf{Description}\\
%  \hline
%  Light Sensor &  \tabincell{l}{Get the current environment brightness: if the brightness is very low, it means the phone is probably \\in pockets/bags or the victim has turned off the room light at night.}  \\
%  \hline
%  Accelerometer & \tabincell{l}{Check the phone is in static status or moving status: if static status, it means the victim is probably \\not using the phone.}\\
%  \hline
%  \tabincell{c}{SDK: \texttt{PowerManager}} & \tabincell{l}{Check the screen is on or off: if the screen is off, it means the victim is probably not using the phone.}\\
%   \hline
%  \tabincell{c}{SDK: \texttt{SimpleDateFormat}} & \tabincell{l}{Get the current system time: for example, 3 AM means the victim is likely to be in deep sleep.}\\
%   \hline
%  Read \texttt{/proc/stat} & \tabincell{l}{Analyze CPU workload: if the workload is low, it means the victim is probably not using his phone.}\\
%   \hline
%   Read \texttt{/proc/meminfo} & \tabincell{l}{Analyze RAM usage: if the usage is low, it means the victim is probably not using his phone.}\\
%   \hline
%\end{tabular}
%\label{tab:context_aware}
%\end{table*}

\subsection{Sound Volume and Sound Pressure Level Experiments}
\label{subsec:cound_experiment}
One issue we concerned is what the sound volume (\texttt{STREAM\_\\MUSIC}) should be set for playing voice commands. The voice of that volume level should be recognized by Google Voice Search and also may not be noticed by the victim. We call it the minimal available sound volume (\emph{MASV} for short). In this section, we only consider the sound volume setting of the speaker (\texttt{MODE\_NORMAL}) in Extended Attack, not the headset (\texttt{MODE\_IN\_CALL}). In Remote Voice Control Attack, the attacker can adjust his own speaking volume directly.

To different models of Android phones, the actual sound pressure and the location of speakers are quite different. Mixed with background white noise, it is impossible to give a fixed sound volume setting. One solution is using the method of successive approximation to pretest. Google Speech-to-Text (STT) is a system built-in service for speech recognition, which uses the same engine as Google Voice Search and can be invoked through \texttt{ACTION\_RECOGNIZE\_SPEECH}~\cite{AndroidRecognizerIntent}. The key point is the app using STT service can get the text content of input voice directly. More details, VoicEmployer sets the sound volume to 1 (the minimum) first, then invokes SST service and plays a prepared test audio file (the content could be "\emph{This is a test message}"). Recognized texts will be returned to VoicEmployer and be compared with the correct texts. If the two sets of texts are the same, VoicEmployer sets system sound volume to 1 and launches GVS-Attack. If the two sets of texts are different, VoicEmployer adjusts the volume to 2 and carries out the same test again. This process will be repeated until finding MASV.

We carried out MASV test experiments in a quiet meeting room of about 8 $m^2$. The environment background sound pressure levels (SPL)~\cite{sound_pressure} was about 48 $dB$. Another relationship we concerned is SPL vs. distance. In the same experiment environment, we recorded the transient peak SPL in different distances (0.5 $m$, 1 $m$ and 2 $m$) from the phone using the above MASV value. Test phones were put on a flat table face up without shelter. Table~\ref{tab:distance_volume1} shows corresponding experiment results. We can find that when the distance exceeds 1 $m$, the SPL is quite low -- only 8 $dB$ higher than the background SPL. Also the SPL would be less than 55 $dB$, which is the general noise level threshold of nocturnal awakenings\footnote{Actually the noise-induced nocturnal awakening is affected by many factors, such as sleep stages, age, gender, smoking, etc. Also sensitivity to noise may vary greatly from one individual to another. These topics are beyond the scope of this paper. More details see the guideline document~\cite{hurtley2009night} of World Health Organization (WHO).  }~\cite{muzet2007environmental}.

\vspace{-5pt}
\begin{table}[htb]
  \caption{Quiet Meeting Room, SPL vs. Distance, unit: $dB$}
  \vspace{5pt}
  \centering
  \begin{tabular}{|lc|c|c|c|}
   \hline
     \textbf{Phone Model} && 0.5 $m$ & 1 $m$ & 2 $m$\\
  \hline
  Samsung Galaxy S3&-- volume 6/15 & 57 & 56 & 54\\
  \hline
  Meizu MX2 &-- volume 5/15 & 58& 54& 53\\
  \hline
  Motolora A953&-- volume 6/15 & 58 & 56 & 54\\
 \hline
\end{tabular}
\label{tab:distance_volume1}
\end{table}

\section{Defense}
\label{sec:defense}
There exist some possible (but not perfect) schemes to defend GVS-Attack. The vulnerability of status checking in Google Search app should be fixed first. When Google Search app receives an \texttt{ACTION\_VOICE\_COMMAND} based Intent, if the phone is securely locked, it must strictly check whether a Bluetooth headset is connected with the phone. If the connect status is true, Voice Dialer can be started. Otherwise, a warning should be provided, like ``\emph{please unlock the device first}''. Therefore, in the situation of secure screen lock, GVS-Attack cannot be launched and the phone is safe.

\emph{From the aspect of app development}, one solution is at run time, Google Voice Search must check the status of speaker in real time, not just in the moment of initialization. If some app is accessing the speaker, Google Voice Search should suspend that app immediately. But this solution may affect other apps' user experience. For example, an IM app may not play its notification sound when new message reaching.

\emph{From the aspect of Android Intent mechanism}, system built-in apps / services should also add customized permissions in the manifest file. if a third-party app wants to invoke a system built-in app / service, it must declare these requirements before installing. So the user needs to confirm the authorization or not installs this app. But this defense method may result in similar abuse problems like system permissions.

\emph{From the aspect of user identification}, speaker recognition (or so called voiceprint recognition) techniques~\cite{beigi2011fundamentals} should be deployed to verify the identity of the speaker. If voiceprint authentication fails, Google Voice Search will not accept the next voice commands. Touchless Control~\cite{touchlesscontrol_url} (a variant of Google Voice Search, only for Motorola phones) has added this function to authenticate the current user through speaking "\emph{OK, Google}"~\cite{moto_gvs}. Also potential problems may be how to defend voice replay attacks.

\section{Discussion}
\label{sec:discussion}
\textbf{Soundless Attack.} One potential limitation of GVS-Attack is that the victim may notice the voice command played by VoicEmployer and interrupt it. One direct idea is whether soundless GVS-Attack could be launched. That is, VoicEmployer imports an audio file to the microphone directly without playing it, like creating a loopback. But in Android OS, the audio recording method is synchronized. It means the microphone cannot support multi-progress accessing at the same time. So when Google Voice Search is accessing microphone to accept voice commands, VoicEmployer cannot access microphone at the same time at least on SDK API level. Also about audio files as the input of microphone, this feature needs the support of kernel / drivers. On Windows and Linux platforms, there exist such drivers (such as VB-Audio Virtual Cable~\cite{vbaudio_url}) to simulate a virtual microphone device. But on Android platform, an app cannot modify kernel or install drivers directly. One feasible solution is to prepare a customized Android version with modified audio drivers. One recent research~\cite{zhou2014peril} has been aware of security risks in Android device driver customizations. Android inherits the driver management methods of Linux and devices are placed under \texttt{/dev} (or \texttt{/sys}) as files. Zhou et al. found the vulnerability that certain important devices become unprotected (permission setting) during a customization. An unauthorized app can get access to sensitive devices, namely user data. Based on this method, similar vulnerability may occur on audio drivers (\texttt{/dev/snd}), but we have not found such a case of unprotected writing privileges on our test devices.

Another perspective of soundless attacks is high-frequency sound (such as higher than 20 kHz), which could be played by the phone and is difficult to hear by humans~\cite{rosen2011signals}. Unfortunately (fortunately for security), Google Voice Search only accepts reasonable human sound frequency range and filters out other ranges. So the idea of high-frequency sound is not impracticable.

Since Google Voice Search is an Internet based service, we also tried to analyze the feasibility of connection hijacking and data package tampering. But after tests, the difficulties lay on the connection is TLS protected and the voice data transmission uses an unclear compressing coding algorithm.

\vspace{5pt}
\textbf{Quiet Vs Noisy.} In GVS-Attack, the attack scenario is quiet environments. The sound volume of played voice commands could be very low and still be recognized by Google Voice Search. Actually we also tested the performance of attacks in noisy environments, such as on the subway and in the canteen. The expected result would be changed, namely the sound volume of played voice commands could be very loud and hides in the background noisy. But test results showed the background noisy (especially human voice) affected the accuracy of speech recognition, to some degree. Also context-aware analysis will become complex and may need the \texttt{RECORD\_AUDIO} permission.

\vspace{5pt}
\textbf{Attack Scope.} For Google Services Framework is pre-installed on nearly all brands of Android devices. Therefore most Android devices can be affected by GVS-Attack, especially these ones equipped with Android 4.1 or upper versions. To other voice assistant apps using Google Speech-to-Text (STT) service, similar attacks could be launched. One example is Speaktoit Assistant\footnote{http://www.speaktoit.com/} (Android version), which has over 5 million installations on Google Play. Other voice assistant apps using independent speech recognition engines also should be reviewed carefully, such as Samsung S Voice app. Also similar attacks may also occur on iOS and Windows Phone platforms. But for the lack of experimental devices, we didn't test them temporarily and left for future research.

\section{Related Work}
\label{sec:relatedWork}
\textbf{Sensor based Attacks.} On mobile platforms, sensor based attacks have been designed and analyzed in several previous papers~\cite{schlegel2011soundcomber, owusu2012accessory, cai2011touchlogger, aviv2012practicality, simon2013pin, placeraider, hasan2013sensing}. One typical example is Soundcomber~\cite{schlegel2011soundcomber}.  Schlegel et al. designed a Trojan with few and innocuous permissions, that can extract targeted private information (such as credit card and PIN number) from the audio sensor of the phone. Also it proved that smartphone based malware can easily achieve targeted, context-aware information discovery from sound recordings. In another research project, through completely opportunistic use of the phone's camera and other sensors, PlaceRaider~\cite{placeraider} can construct three dimensional models of indoor environments and steal virtual objects.

The work of~\cite{owusu2012accessory} showed that accelerometer readings are a powerful side channel that can be used to extract entire sequences of entered text on a smartphone touchscreen keyboard. In~\cite{simon2013pin}, the designed side-channel attack utilizes the video camera and microphone to infer PINs entered on a number-only soft keyboard on a smartphone. The microphone is used to detect touch events, while the camera is used to estimate the smartphone's orientation, and correlate it to the position of the digit tapped by the user. The work of~\cite{hasan2013sensing} studied environmental sensor-based covert channels in mobile malware. Out-of-band command and control channels could be based on acoustic, light, magnetic and vibrational signaling. This research is a bit like our GVS-Attack in the aspect of voice command transmission. The differences are that, in GVS-Attack, the command receiver (Google Voice Search) is not a part of malware (VoicEmployer) and the \texttt{RECORD\_AUDIO} permission is not necessary. So our attack scheme is more insidious.

Sensors also could be used for fingerprinting devices. A mechanism was proposed by~\cite{dey2014accelprint} that smartphone accelerometers possess unique fingerprints, which can be exploited for tracking users. A similar fingerprinting method was designed with microphone and speaker in~\cite{das2014poster}. Through playback and recording of audio samples, this method could uniquely identify an individual device based on sound analysis.

\vspace{5pt}
\textbf{Inter-Application Communication.} In~\cite{chin2011analyzing}, Chin et al. focused on Intent-based attack surfaces. It analyzed unauthorized Intent receipt can leak user information. Data can be stolen by eavesdroppers and permissions can be accidentally transferred between apps. Another attack type is Intent spoofing, that a malicious app sends an Intent to an exported component. If the victim app takes some action upon receipt of such an Intent, the attack can trigger that action. In their following work~\cite{kantola2012reducing},
Kantola et al. proposed modifications to the Android platform to detect and protect inter-application messages that should have been intra-application messages. The target is to automatically reduce attack surfaces in legacy apps.

\vspace{5pt}
\textbf{Application Analysis} Android permission-based security has been analyzed from many aspects. For example, permission specification and least-privilege security problems were studied in~\cite{felt2011android} and~\cite{au2012pscout}. Both of them designed corresponding static analysis tools to detect over-privilege problems. Permission escalation and leakage problems are also hot research topics. Related research include~\cite{felt2011permission, davi2011privilege, bugiel2012towards, chan2012droidchecker, grace2012systematic}. Permission-based behavioral footprinting was used to detect known malwares in Android market at large-scale in~\cite{zhou2012hey}. The work of~\cite{wu2013impact} noticed the problem of pre-installed apps. Wu et al. found a lot of those apps were overly privileged for vendor customizations.

To system code level analysis, dynamic analysis technique is an efficient solution. TaintDroid~\cite{enck2010taintdroid} is a system-wide dynamic taint tracking and analysis system capable of simultaneously tracking multiple sources of sensitive data. It automatically labels (taints) data from privacy-sensitive sources and transitively applies labels as sensitive data propagates through program variables, files, and interprocess messages. DroidScope~\cite{yan2012droidscope} is an emulation based Android malware analysis engine that can be used to analyze the Java and native components of Android apps. Unlike current desktop malware analysis platforms, DroidScope reconstructs both the OS-level and Java-level semantics simultaneously and seamlessly.

But these analysis methods are useless to our GVS-Attack, because VoicEmployer doesn't touch sensitive data and execute sensitive operations directly. The voice commands are transferred out of the phone, that is, an uncontrolled physical communication channel. Also the remote data transmission function is based on the call channel and the transmission media is sound. These brand-new features break previous checking and protecting mechanism.

%Google Voice Search vulnerability on chrome reported.

% vim: tw=0

\section{Conclusion}
\label{sec:conclusion}
This paper proposes a novel permission bypassing attack method based on Android system build-in voice assistant module - Google Voice Search and the speaker. Also the app launching attacks doesn't require any permission. But achieved malicious targets could be quite dangerous and practical, from privacy stealing to remote voice control. Also utilizing a vulnerability found by us in Google Search app, this attack can dial arbitrary malicious numbers even when the phone is securely locked. Related in-depth topics are also discussed, including context-aware analysis, minimal available sound volume, soundless attack, etc. Through experiments, the feasibility of our attack schemes has been demonstrated in real world. This research may inspire application developers and researchers rethink that zero permission doesn't mean safety and the speaker can be treated as a new attack surface.

%\vspace{10pt}

%\flushend
% If you are using bibtex:

%\bibliographystyle{plain}
\bibliographystyle{abbrv}
%\bibliographystyle{IEEEtran}
%\balance
\bibliography{refs}

\appendix
%Appendix A
\section{Permissions Statistics}
%\subsection{Google Search App}
According to Android API Level 19, Google Search app (version 3.4.16.1149292.arm) declares / possesses the following system defined permissions:
 \begin{enumerate}
   \item  \texttt{ACCESS\_COARSE\_LOCATION}
   \item  \texttt{ACCESS\_FINE\_LOCATION}
   \item  \texttt{ACCESS\_NETWORK\_STATE}
   \item  \texttt{ACCESS\_WIFI\_STATE}
   \item  \texttt{BIND\_APPWIDGET}
   \item  \texttt{BLUETOOTH}
   \item  \texttt{BROADCAST\_STICKY}
   \item  \texttt{CALL\_PHONE}
   \item  \texttt{GET\_ACCOUNTS}
   \item  \texttt{GLOBAL\_SEARCH}
   \item  \texttt{INTERNET}
   \item  \texttt{MANAGE\_ACCOUNTS}
   \item  \texttt{MEDIA\_CONTENT\_CONTROL}
   \item  \texttt{MODIFY\_AUDIO\_SETTINGS}
   \item  \texttt{READ\_CALENDAR}
   \item  \texttt{READ\_CONTACTS}
   \item  \texttt{READ\_EXTERNAL\_STORAGE}
   \item  \texttt{READ\_HISTORY\_BOOKMARKS}
   \item  \texttt{READ\_PROFILE}
   \item  \texttt{READ\_SMS}
   \item  \texttt{READ\_SYNC\_SETTINGS}
   \item  \texttt{RECEIVE\_BOOT\_COMPLETED}
   \item  \texttt{RECORD\_AUDIO}
   \item  \texttt{SEND\_SMS}
   \item  \texttt{SET\_ALARM}
   \item  \texttt{SET\_WALLPAPER}
   \item  \texttt{SET\_WALLPAPER\_HINTS}
   \item  \texttt{STATUS\_BAR}
   \item  \texttt{USE\_CREDENTIALS}
   \item  \texttt{VIBRATE}
   \item  \texttt{WAKE\_LOCK}
   \item  \texttt{WRITE\_CALENDAR}
   \item  \texttt{WRITE\_EXTERNAL\_STORAGE}
   \item  \texttt{WRITE\_SETTINGS}
   \item  \texttt{WRITE\_SMS}
 \end{enumerate}

 Non-public permissions:

 \begin{enumerate}
   \item \texttt{CAPTURE\_AUDIO\_HOTWORD}
   \item \texttt{STOP\_APP\_SWITCHES}
   \item \texttt{PRELOAD}
 \end{enumerate}

\end{document}